\documentclass[acus]{JAC2003}

\usepackage{graphicx}
\usepackage{booktabs}
\usepackage{tabularx}

\begin{document}
\title{SYNCHRONOUS PHASE SHIFT AT LHC}

\author{J.~F.~Esteban M\"uller\thanks{jesteban@cern.ch}, P.~Baudrenghien, G.~Iadarola, T.~Mastoridis, G.~Papotti,\\ G.~Rumolo, E.~Shaposhnikova, D.~Valuch, CERN, Geneva, Switzerland}

\maketitle

\begin{abstract}
   The electron cloud in vacuum pipes of accelerators of positively charged particle beams causes a beam energy loss which could be estimated from the synchronous phase. Measurements done with beams of 75~ns, 50~ns,  and 25~ns bunch spacing in the LHC for some fills in 2010 and 2011 show that the average energy loss depends on the total beam intensity in the ring. Later measurements during the scrubbing run with 50~ns beams show the reduction of the electron cloud due to scrubbing. Finally, measurements of the individual bunch phase give us information about the electron cloud build-up inside the batch and from batch to batch.
\end{abstract}

\section{INTRODUCTION}
The beam circulating in the ring can lose energy due to various effects, among which we can highlight as the most important: synchrotron radiation, the resistive impedance of the accelerator, and the interaction between the beam and the electron cloud. This energy loss is compensated by the RF system by synchronous phase shift. To achieve this, in the absence of acceleration, the synchronous phase shift $\phi_s $ from the stable phase ($\pi$ for the LHC being above transition) should be:
\begin{equation}
\sin \phi_s = \frac{W}{e V},
\end{equation}
where $W$ is the energy loss per turn and per particle and $V$ is the RF voltage amplitude. The synchronous phase can therefore be used to calculate the bunch energy loss and the total energy loss of the beam.

The energy loss per particle by synchrotron radiation does not depend on the total intensity, but on the energy of the particle and its bending radius, so it gives a constant phase offset for all bunches. In the case of the energy loss per particle due to broad-band resistive impedance, it depends on the bunch intensity and length. If bunch intensities and lengths are similar for all bunches and change only slightly during the fill, a phase offset is also similar for all bunches but it can be different from fill to fill depending on the beam parameters. Finally, the energy loss due to electron cloud depends on bunch intensities and lengths, bunch and batch spacings, and the electron cloud generated by the passage of the previous bunches as well as uncaptured beam. The synchronous phase can be therefore used to calculate the energy loss due to electron cloud if the offset given by the other factors is removed~\cite{elena_slides}.

\section{EXPERIMENTAL CONDITIONS}
To separate the synchronous phase shift due to electron cloud from beam loading effects, the Beam Phase Module~\cite{phase_module} from the LHC Low Level RF was used. This module measures the synchronous phase as the difference in phase between the beam and the vector sum of the voltage in the 8 cavities. This signal is called phase error and is used by the Phase Loop to damp coherent dipole mode synchrotron oscillations.

\begin{figure}[htb]
 \centering
 \includegraphics[width=\columnwidth]{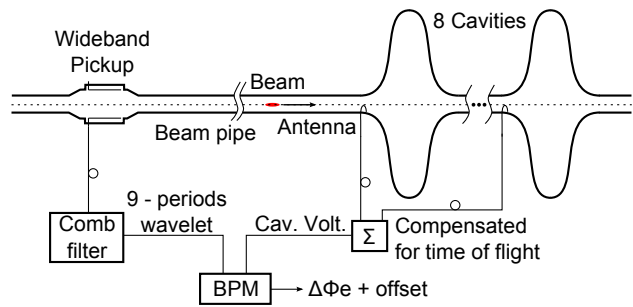}
 \caption{Simplified scheme of the phase error measurement. A wavelet is generated from the wideband pickup signal and is compared in phase with the vector sum of the 8 cavities voltages in the Beam Phase Module (BPM). \label{fig:meas_scheme}}
\end{figure}

The measurement scheme is shown in Fig.~\ref{fig:meas_scheme}. The signal from a 3~GHz band-width pickup is fed into a strip-line comb filter transforming a single (bunch) pulse into a wavelet at 400.8 MHz lasting for 9 RF periods. The output is the beam signal input to the Beam Phase Module. The second RF input is the vector sum of the eight cavity antenna signals with delays to compensate the time of flight between cavities. Two analog in-phase/quadrature (I/Q) demodulators transform the beam signal and cavity sum into (I,Q) pairs with respect to the Local Oscillator (LO) (400.8 MHz). Intermediate Frequency (IF) signals are bunch synchronously sampled at 40.08 MHz (bunch frequency). An FPGA computes the phase and amplitude of the 400.8 MHz component of each individual bunch with respect to the cavity sum. The precision is mainly limited by thermal drifts. The average phase error of the first batch (consisting of 12 bunches) was used as a reference to remove the offset.

During 2011 operation, modifications were done in the Beam Phase Module, including some changes in parameters and hardware. Changes in parameters affect the offset level, which is not important for the measurements because it is removed. The hardware modifications done on $20^\textrm{th}$ May 2011 improved the quality of the signals.

In the first part of this Note, we describe measurements of the average phase error for all the bunches in the ring, which were used to observe the effect of the electron cloud on the beam as a function of the total intensity and to estimate the average energy loss per particle. In the second part of the Note, we present measurements of the individual bunch phase error, which provides information about the electron cloud build up and more accurate estimations of the total energy loss.

\section{MEASUREMENTS OF THE AVERAGE PHASE ERROR}
The measured average synchronous phase of all the bunches in the ring is:
\begin{equation}
 \label{eq:av_ph}
 \langle \phi_s \rangle = \frac{1}{\displaystyle K} \displaystyle \sum^{K}_{k=1} \phi_{sk},
\end{equation}
where $K$ is the number of bunches and $\phi_{sk}$ is the synchronous phase of bunch $k$. For results presented in this section the average phase error signal from the Phase Module was used as a measurement of the average synchronous phase.

The average energy loss per particle and per turn can be found as:

\begin{equation}
\label{eq:av_energy_loss}
\langle U \rangle = \frac{\displaystyle e \, V \, \sum_{k=1}^{K} N_{k} \, \sin \phi_{sk}}{\displaystyle \sum_{k=1}^{K} N_{k}},
\end{equation}
where $N_{k}$ is the intensity of the bunch $k$.

For small shifts ($\phi_{sk}\ll1$) and similar bunch intensities ($N_{k}\approx N_b, \forall k$), Eq.~(\ref{eq:av_energy_loss}) becomes:

\begin{equation}
\label{eq:app_energy_loss}
\langle U \rangle \approx e \, V \, \langle \phi_{s} \rangle ,
\end{equation}
and the total power loss of the beam is:
\begin{equation}
\label{eq:power_loss}
P_L \approx K \,N_b \, e \, V \, f_{rev} \, \langle \phi_{s} \rangle ,
\end{equation}
where $K \, N_b$ is the total intensity and $f_{rev}$ is the revolution frequency.  Eq.~(\ref{eq:app_energy_loss}) gives the relation between the beam energy loss and the average phase error.

\begin{figure}[htb]
 \centering
 \includegraphics[width=\columnwidth]{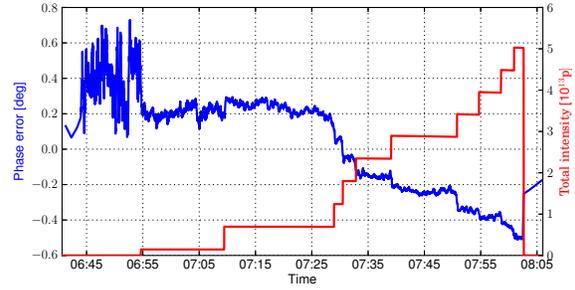}
 \caption{Average phase error (blue) and total intensity (red). Note the phase shifts after each batch injection. Beam 1.  Fill 1502 (50~ns, 20-11-2010). \label{fig:example1}}
\end{figure}

\begin{figure}[htb]
 \centering
 \includegraphics[width=\columnwidth]{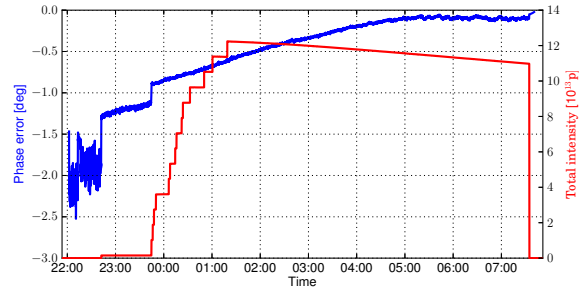}
 \caption{Average phase error (blue) and total intensity (red). Note the drift of the phase error due to thermal effects.  Beam 2. Fill 1694 (50~ns,  10-04-2010). \label{fig:example2}}
\end{figure}

The average phase error signal was extracted from the Timber database, where the Beam Phase Module stores it every second. A typical example of the average phase error signal for Beam~1 is shown in  Fig.~\ref{fig:example1}. It can be observed after the second batch injection that the phase error shifts at each batch injection due to the increase of the electron cloud density. For Beam~2, we systematically observe a drift in the phase error probably caused by thermal effects and much larger than the effect expected from electron cloud, see Fig.~\ref{fig:example2}. Consequently, the observations presented in this section are for Beam~1 only. There are also shifts and drifts in the phase error for some fills that are not understood yet. We have ignored these fills and we present in Table~\ref{tab:fills} the fills selected for the measurements.

\begin{table*}[tb]
 \centering
 \newcolumntype{C}{>{\centering\arraybackslash}X}%
 \caption{Fills used for the phase shift measurements.\label{tab:fills}}
 \begin{tabularx}{0.8\textwidth}{ C C C C C C C }
  Year\textbackslash Spacing & \multicolumn{2}{c}{75~ns}  &  \multicolumn{2}{c}{50~ns} & \multicolumn{2}{c}{25~ns} \\
 \toprule
   	& Fill		& Date	& Fill		& Date	& Fill	& Date \\
 \toprule
 2010	& 1497	& 17-11	& 1500	& 19-11	& \multicolumn{2}{c}{--} \\
 	& 1498	& 18-11	& 1502	& 20-11	& \multicolumn{2}{c}{ }\\
	& 1499	& 19-11	&		&		& & \\
 \midrule
 2011 	& 1645 	& 22-03	& 1675	& 06-04	& 1905	& 20-06 \\
  	& 1647 	& 23-03	& 1677	& 06-04	& 2185	& 07-10 \\
 	& 1671 	& 04-04	& 1683	& 09-04	& 2212	& 14-10 \\
 	&		&		& 1685	& 09-04	& 2213	& 14-10 \\
	&		&		& 1686	& 10-04	& 2214	& 14-10\\
	&		&		& 1687	& 10-04	& 2250	& 24-10 \\
	&		&		& 1688	& 10-04	& 2251	& 25-10 \\
	&		&		& 1689	& 10-04	& 		& \\
	&		&		& 1690	& 10-04	& 		& \\
	&		&		& 1692	& 10-04	& 		& \\
	&		&		& 1694	& 10-04	& 		& \\
 \bottomrule
 \end{tabularx}
\end{table*}

The phase error was measured after each injection and represented versus the total intensity at the same time in the plots shown in this section. The phase was averaged during 40 seconds after injection. Then a linear fit is applied to remove the constant offset introduced by the measurement and other factors described in the introduction. In the measurements a decrease in phase error means an increase in energy loss. In all plots in this section the sign has been changed so that an increase in phase error corresponds to an increase in energy loss.

It is important to mention that the voltage program was changed in 2011 with respect to 2010. In 2010 the voltage at the flat bottom was 3.5~MV, while it was set up to 6~MV in 2011. It means that for the same energy loss, the synchronous phase in 2010 should be almost twice larger than in 2011.

\subsection{Observations from 2010}
\begin{figure}[htb]
 \centering
 \includegraphics[width=\columnwidth]{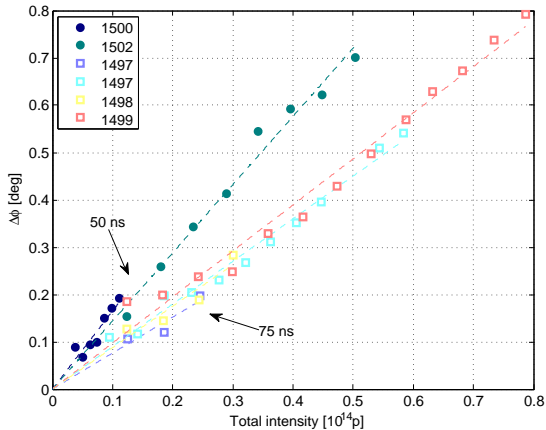}
 \caption{Average phase error at injections as a function of the total intensity in the ring for fills in 2010. Higher energy loss for the 50~ns beam (circles) than for the 75~ns beam (squares). V~=~3.5~MV.\label{fig:2010}}
\end{figure}

During 2010, there were fills with 150~ns, 75~ns, and 50~ns bunch spacings. The phase error data were available from November and they include only a few fills of 75~ns beam and some measurements during the setting-up of the 50~ns beam. The measurements of the phase error as a function of the total intensity in the ring are shown in Fig.~\ref{fig:2010}. As can be seen the phase error increases almost linearly with the total beam intensity. That means that for these beams the electron cloud density was increasing with the intensity. Note also that the phase shift for the 50~ns beam has a steeper slope than for the 75~ns beam, which implies that the electron cloud generation depends on the bunch spacing, and it is higher for smaller bunch spacings.

\subsection{Observations from 2011}
\begin{figure}[htb]
 \centering
 \includegraphics[width=\columnwidth]{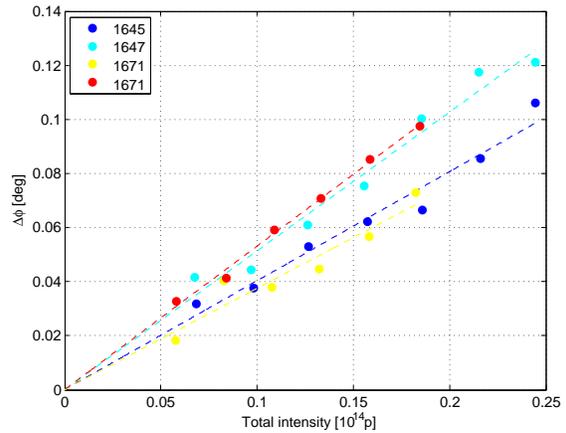}
 \caption{Average phase error at injections as a function of the total intensity in the ring. 75~ns beam. 2011. V~=~6~MV. \label{fig:2011-75}}
\end{figure}

In 2011 the LHC had fills with 75~ns, 50~ns, and 25~ns beams. The 75~ns beam was used at the beginning of the year and the observed electron cloud effect  was negligible. The measurement results are shown in Fig.~\ref{fig:2011-75}.

\begin{figure}[htb]
 \centering
 \includegraphics[width=\columnwidth]{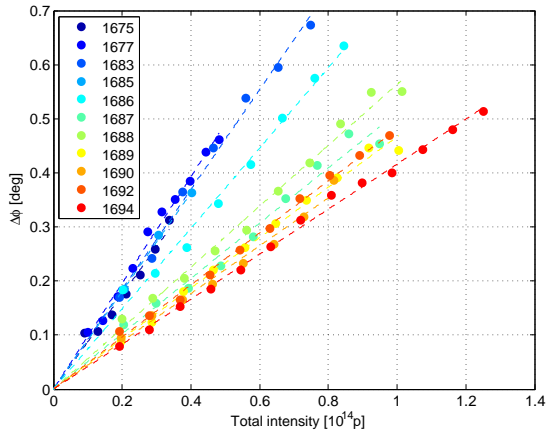}
 \caption{Average phase error at injections as a function of the total intensity in the ring for the 50~ns beam during the scrubbing run. Note the decrease in the slope of the curves between the beginning and the end of the scrubbing run. 2011. V~=~6~MV. \label{fig:2011-50}}
\end{figure}

\begin{figure}[htb]
 \centering
 \includegraphics[width=\columnwidth]{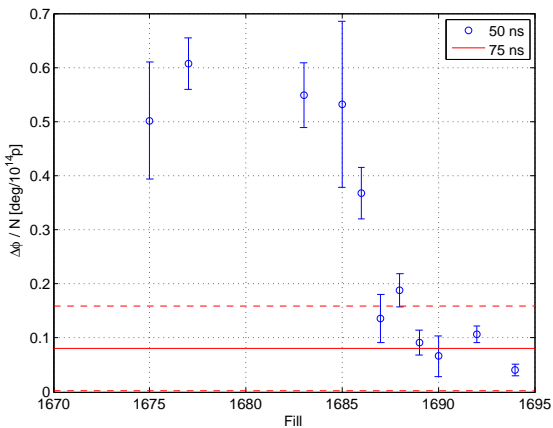}
 \caption{ In blue circles, ratio of the phase shift to total intensity for the 50~ns beam. In red lines, mean value (solid) and standard deviation (dashed) of the same ratio for the 75~ns beam. Note that after scrubbing run the value of the ratio for the 50~ns beam converged to the 75~ns beam value. 2011. V~=~6~MV. \label{fig:2011-50-75}}
\end{figure}

Then a scrubbing run was done with the 50~ns beam in April. Figure~\ref{fig:2011-50} shows most of these fills and it is apparent that the slope of the curves is decreasing from fill to fill. This is an evidence that the electron cloud density was decreasing due to a reduction of the Secondary Emission Yield (SEY) and a proof that the scrubbing run was beneficial. A comparison of the slopes for the 50~ns beam during the scrubbing run and the 75~ns beam is shown in Fig.~\ref{fig:2011-50-75}. At the beginning of the 2011 run, the slope was much higher for the 50~ns beam than for the 75~ns beam, but at the end they were similar. The electron cloud reduction was also seen by other measurements, such as the pressure rise, heat load in the arcs, and emittance growth~\cite{lhc-ecloud-giovanni}.

\begin{figure}[htb]
 \centering
 \includegraphics[width=\columnwidth]{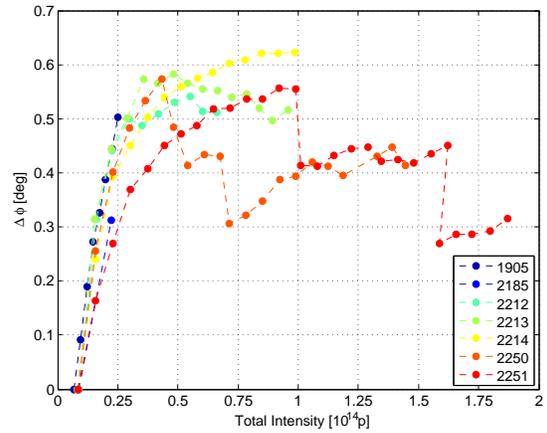}
 \caption{Average phase error at injections as a function of the total intensity in the ring for the 25~ns beam. The saturation effect of the electron cloud is visible after the injection of few batches. Note that the time between injections is not constant. If that time is too long, it leads to an electron cloud density decrease due to intensity loss. 2011. V~=~6~MV. \label{fig:2011-25}}
\end{figure}

During 2011 there were some Machine Development studies (MD) with 25~ns beam. The measurements done during these MDs sessions are shown in Fig.~\ref{fig:2011-25}. For the 25~ns beam the electron cloud density reached saturation after some injections, as can be seen in Fig.~\ref{fig:2011-25}. The effect was so strong that the last bunches of each batch lost continuously intensity during the fills, which reduced the electron cloud density. This is also visible in Fig.~\ref{fig:2011-25} as a decrease of phase error when the time between injections is long enough.

\begin{figure}[htb]
 \centering
 \includegraphics[width=\columnwidth]{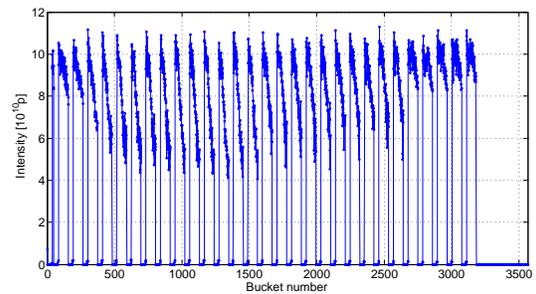}
 \caption{Typical bunch intensity pattern for the 25~ns beam, with a triangular batch shape due to particle loss from the electron cloud effect. The last five batches were injected later and they were still losing particles. Beam 1. Fill~2251 (25~ns, 25-10-2011). \label{fig:25ns_bunch_pattern}}
\end{figure}

The typical bunch intensity pattern for the 25~ns beam can be seen in Fig.~\ref{fig:25ns_bunch_pattern}, where one can observe the triangular shape of the batches due to intensity loss previously mentioned. The difference in the intensity of the bunches implies that the average energy loss is not directly proportional to the average phase error for this case, since the approximations done to derive Eq.~(\ref{eq:app_energy_loss}) are not valid anymore. In addition, the phase shift due to the broad-band impedance depends on the bunch intensities and lengths. Therefore it is different from bunch to bunch and affects the average phase error. These effects were not taken into account in this Note; however, for the first injections, while all the bunches still have similar intensities, a scrubbing effect is visible between the first and the last 25~ns fill.

\subsection{Power Loss Estimation}
\begin{figure}[htb]
 \centering
 \includegraphics[width=\columnwidth]{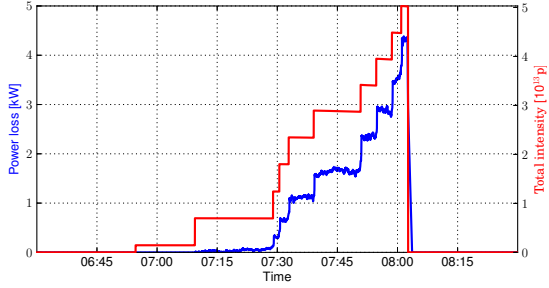}
 \caption{ Total beam power loss (blue) and total intensity (red). Beam 1. Fill 1502 (50~ns,  20-11-2010). \label{fig:power_loss}}
\end{figure}

The total beam power loss for the same 50~ns beam fill as in Fig.~\ref{fig:example1} example is shown in Fig.~\ref{fig:power_loss}, calculated using Eq.~(\ref{eq:power_loss}). It can be seen that the steps in the power loss are bigger after each injection, as the power loss per particle increases with the total intensity in the ring. The maximum power loss is 4.35~kW for $5 \times 10^{13}$ particles in the ring.

\section{BUNCH BY BUNCH MEASUREMENTS}
Measurements of the bunch by bunch phase error give information about the electron cloud build-up. In this case the energy loss per turn for the $k^\textrm{th}$ bunch, for $\phi_{sk} \ll 1$, is:
\begin{equation}
\label{eq:bunch_energy_loss}
U_k =N_k \,  e \, V \, \phi_{sk},
\end{equation}
and the power loss per bunch is:
\begin{equation}
\label{eq:bunch_power_loss}
P_{Lk} = N_k \,  e \, V \, f_{rev} \, \phi_{sk}
\end{equation}

The bunch by bunch phase error data are not logged automatically in the Timber database. The acquisition of the data was available from April, 2011 and measurements from only a few fills of the 50~ns and 25~ns beams were done, because the acquisition is done manually and has conflicts with the CCC fixed display, which requires the operators assistance.

The system has been upgraded in 2012. The phase measurement boards have been moved into the UX45 Cavern, closer to the pickups and the cavities. That should reduce the sensitivity to thermal fluctuations and the effect of localized mismatches on the RF cables, and it will provide more accurate measurements.

\subsection{Measurements Correction}
The signal at the Beam Phase Module following the passage of a single bunch in the pick-up should be strictly limited to a 20~ns long wavelet at 400~MHz in order to accurately measure  the phase of each bunch independently. Imperfections in the Beam Phase Module response, reflections at its input, and localized mismatch in the cables distort the single bunch measurement by adding contributions of other bunches. Assuming linearity, this effect can be compensated by deconvolving with the impulse response.


\begin{figure}[htb]
 \centering
 \includegraphics[width=\columnwidth]{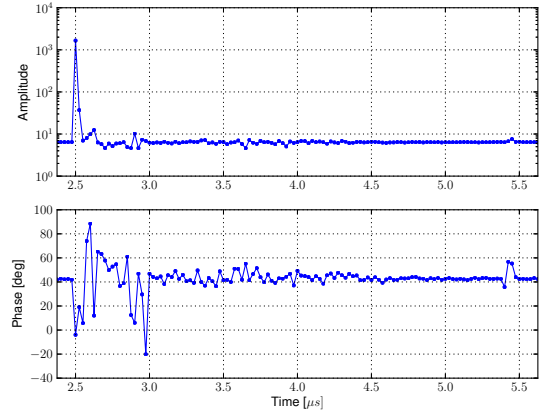}
 \caption{Amplitude (top) and phase (bottom) of the pick-up signal measured after a single bunch passage. The bunch passes at $2.5 \mu s$. Beam 1. Fill~2273 (Single bunch, 04-11-2011). \label{fig:SB_response_B1}}
\end{figure}

\begin{figure}[htb]
 \centering
 \includegraphics[width=\columnwidth]{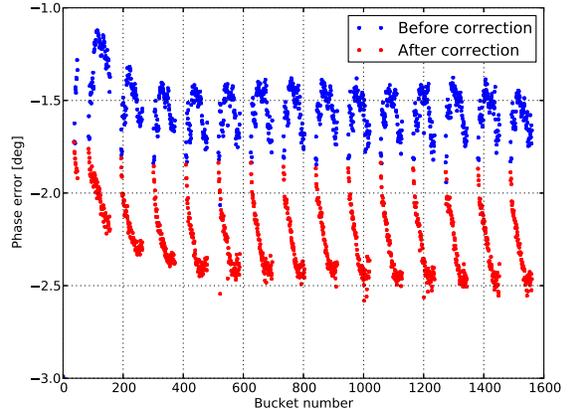}
 \caption{Measured bunch by bunch phase error before (blue) and after (red) correction. Beam~2. Fill~2251 (25~ns, 25-10-2011). \label{fig:BbB_corrected}}
\end{figure}

A measurement with a single bunch of $1.1 \times 10^{11}$ p in the ring was done to extract the impulse response. The measurement is shown in Fig.~\ref{fig:SB_response_B1}. 
The acquired data is post-processed using this impulse response. An example of measurements before and after the processing can be seen in Fig.~\ref{fig:BbB_corrected}.

\begin{figure}[htb]
 \centering
 \includegraphics[width=\columnwidth]{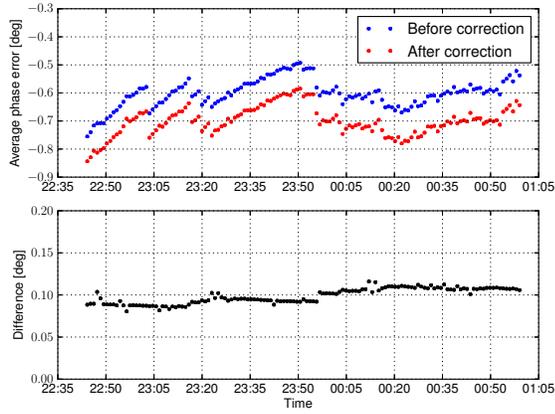}
 \caption{Comparison between average of the bunch by bunch phase error measurements before (blue) and after (red) correction (top) and the difference between both measurements (bottom). Note that the difference is approximately constant. Beam 1. Fill 2250 (25~ns, 24-10-2011). \label{fig:av_ph_correction}}
\end{figure}

The average phase error (used in the first part of this Note) is practically not affected by these corrections, as can be seen in Fig.~\ref{fig:av_ph_correction}, where the average of the bunch by bunch phase error with and without corrections is shown as a function of time for measurements with 25~ns beam. The difference between the measurements before and after corrections is approximately constant and only changes the offset level.

\subsection{Beam Loading}
The phase error measurements are in principle not affected by beam loading. In order to check the independence of the phase error measurements on the beam loading, the phase error was compared during the same fill with One-Turn Feedback switched on and off. The One-Turn Feedback is a system that increases the RF feedback gain at the revolution harmonics. As such, it reduces the beam loading in the cavities. Measurements were done during the commissioning of the One-Turn Feedback with 25~ns spacing beam on $24^\textrm{th}$ October, 2011~\cite{OTFB}.

Measurements of bunch positions were then taken from the Beam Quality Monitor (BQM). The BQM measures longitudinal bunch profiles from a Wall Current Monitor and uses them to calculate longitudinal parameters of the beam~\cite{bqm}. One of them is the bunch position, a measurement of the bunch center position. It measures the absolute distance between successive bunches and therefore includes effects from both transient beam loading and electron cloud. The difference between the bunch position and the phase error is mainly due to beam loading.

\begin{figure}[htb]
 \centering
 \includegraphics[width=\columnwidth]{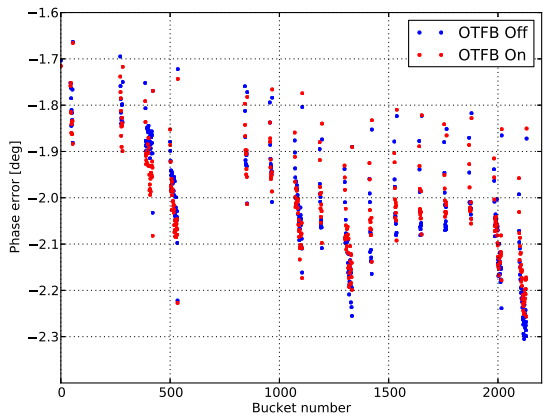}
 \caption{Bunch by bunch phase error with One-Turn Feedback off (blue) and on (red). Note that the signals are very similar. Beam~2. Fill~2248 (25~ns, 24-10-2011). \label{fig:1T-FB_effect_ph_err}}
\end{figure}

\begin{figure}[htb]
 \centering
 \includegraphics[width=\columnwidth]{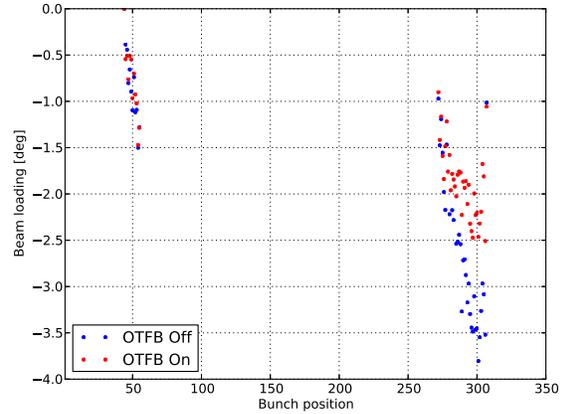}
 \caption{Bunch by bunch phase shift due to beam loading with One-Turn Feedback off~(blue) and on~(red).  Note the reduction of beam loading when the One-Turn Feedback is on. Beam~2. Fill~2248 (25~ns, 24-10-2011). \label{fig:1T-FB_effect_beam_loading}}
\end{figure}

Measurements of the bunch by bunch phase error and phase shift due to beam loading at the same time are shown, respectively, in Fig.~\ref{fig:1T-FB_effect_ph_err} and Fig.~\ref{fig:1T-FB_effect_beam_loading}. While the phase error signals are very similar, there is a clear reduction in the beam loading when the One-Turn Feedback was switched on. We can conclude that the phase error measurements are almost not affected by beam loading.

\subsection{The 50~ns Beam}
\begin{figure}[htb]
 \centering
 \includegraphics[width=\columnwidth]{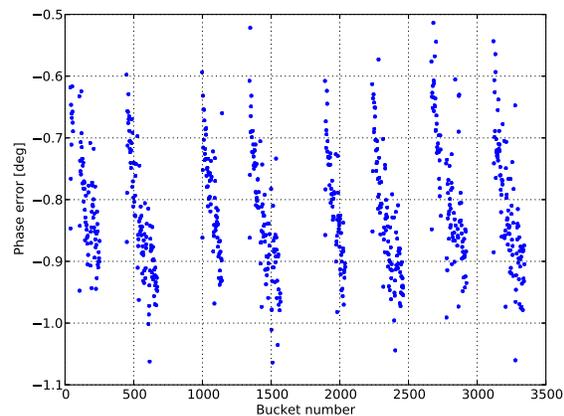}
 \caption{Bunch by bunch corrected phase error. Beam 1. Fill 1798 (50~ns, 21-05-2011). \label{fig:BbB_ph_50ns_1}}
\end{figure}

The measurement for the 50~ns beam from $21^{\textrm{st}}$ May, 2011, is shown in Fig.~\ref{fig:BbB_ph_50ns_1}. Although this measurement was made after the scrubbing run, it is possible to see how the electron cloud is increasing along the batch. The phase error of the bunches at the end of the batches is larger than at the beginning, which means that the particles in the batch tails lose more energy, due to the continuous increase of the electron cloud density. The electron cloud density in this case is small.

\begin{figure}[htb]
 \centering
 \includegraphics[width=\columnwidth]{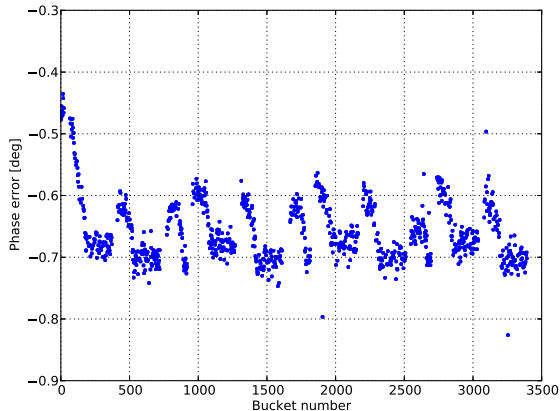}
 \caption{Bunch by bunch corrected phase error. Beam 1. Fill 2267 (50~ns,  30-10-2011). \label{fig:BbB_ph_50ns_2}}
\end{figure}

Fig.~\ref{fig:BbB_ph_50ns_2} shows data from one of the last fills with protons of the LHC in 2011. In this example, the electron cloud is even lower than in the previous one, Fig.~\ref{fig:BbB_ph_50ns_1}, because the machine was scrubbed during the physics fills. The total excursion in phase error reduced from around 0.5 degrees in May to 0.3 degrees in October. In any case, one can consider that the electron cloud is negligible for the 50~ns beam after the scrubbing run.

Another difference between the two examples is that in the first one, Fig.~\ref{fig:BbB_ph_50ns_1}, the spacing between batches is larger than in the example of Fig.~\ref{fig:BbB_ph_50ns_2}. Large spacing permits the electron cloud to decay before the arrival of the next batch, which can also be seen from the fact that all the first bunches of each batch have the same phase error. In the second plot, the electron cloud does not disappear completely before the following batch arrives.

\subsection{The 25~ns Beam}
\begin{figure}[htb]
 \centering
 \includegraphics[width=\columnwidth]{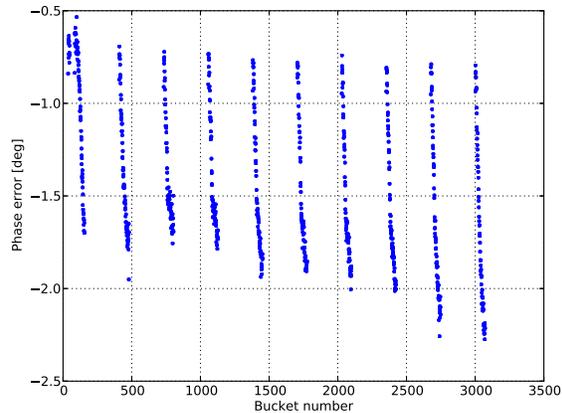}
 \caption{Bunch by bunch corrected phase error for beam with large batch spacing. Beam 1. Fill 2212 (25~ns, 14-10-2011). \label{fig:BbB_ph_25ns_1}}
\end{figure}

\begin{figure}[htb]
 \centering
 \includegraphics[width=\columnwidth]{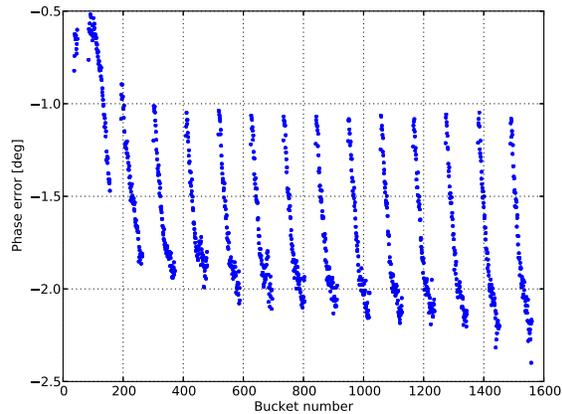}
 \caption{Bunch by bunch corrected phase error. Beam 1. Fill 2214 (25~ns, 14-10-2011). \label{fig:BbB_ph_25ns_2}}
\end{figure}

Measurements shown in Figs.~\ref{fig:BbB_ph_25ns_1} and \ref{fig:BbB_ph_25ns_2} for the 25~ns beam are very similar to the 50~ns beam case. This time both measurements were taken on the same day. In the first plot the batch spacing is larger ($6.325 \, \mu$s) and the electron cloud reduces significantly between batches. In the second figure, there is a residual electron cloud density remaining from the previous batch (the batch spacing was 925 ns) , which allows the electron cloud to grow to a higher level each time. It is visible in Fig.~\ref{fig:BbB_ph_25ns_1} that the phase of the first bunch of each batch is similar, while in Fig.~\ref{fig:BbB_ph_25ns_2} the residual electron cloud from the previous batch shifts the phase of the first bunch of the following batch for the first few batches and then a saturation effect is observed.

\begin{figure}[htb]
 \centering
 \includegraphics[width=\columnwidth]{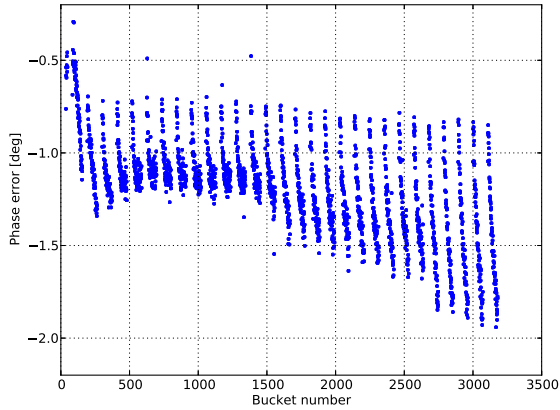}
 \caption{Bunch by bunch corrected phase error. Beam 1. Fill 2251 (25~ns, 25-10-2011) \label{fig:BbB_ph_25ns_3}}
\end{figure}

Finally, the last fill of 2011 with a 25~ns beam can be seen in Fig.~\ref{fig:BbB_ph_25ns_3}. It is clear that the machine is more scrubbed now, because the total excursion in phase error is 1.6 degrees. However, the electron cloud is still significant. It is also interesting to see in this figure that the last batches produce more electron cloud. This is because the first injected batches had been in the machine for a while and their last bunches had lost intensity (due to the electron cloud), as shown in the example in Fig.~\ref{fig:25ns_bunch_pattern}.

\subsection{Power Loss}
\begin{figure}[htb]
 \centering
 \includegraphics[width=\columnwidth]{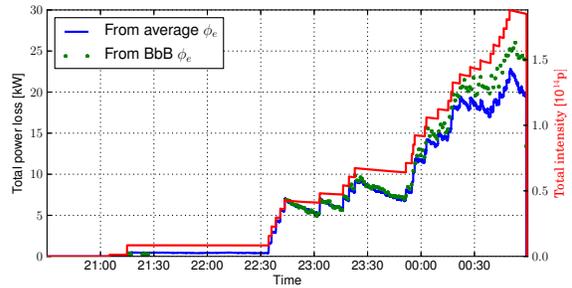}
 \caption{Total beam power loss from average phase error measurements (blue) and from bunch by bunch measurements (green), and total intensity (red). Note that the two measurements of the power loss are not comparable after some injections, when the bunch intensities are different due to the losses caused by the electron cloud. Beam 1. Fill 2250 (25~ns, 24-10-2011). \label{fig:power_loss_av_vs_BbB}}
\end{figure}

The power loss per bunch was calculated using Eq.~(\ref{eq:bunch_power_loss}). It can be used to find the total power loss of the beam more accurately than using the average phase error Eq.~(\ref{eq:power_loss}) if bunch intensities are different, as it was the case for the measurements of the 25~ns beam. An example of the power loss of a 25~ns beam is shown in Fig.~\ref{fig:power_loss_av_vs_BbB}, where the difference between the two methods can be seen, mainly at the end of the fill due to intensity losses. The power loss is approximately a factor of 2 higher than for the 50~ns beam before the scrubbing run (Fig.~\ref{fig:power_loss}) for the same intensity ($5 \times 10^{13}$ p).

\section{CONCLUSIONS}
The measurement of the phase shift has been proven to be a good and novel method to observe and characterize the electron cloud effect. It is possible in the LHC due to the very high accuracy of the phase error measurements. The average phase error gives an estimation of the total beam power loss due to electron cloud and is useful to see the progress in the scrubbing process of the machine. Additionally, the bunch by bunch signal provides more information about the build up of the electron cloud and can be used to calculate the power loss in a more accurate way.

\section{ACKNOWLEDGMENTS}
We wish to thank G.~Arduini, A.~Burov, A. Butterworth, G. Hagmann, W.~Hofle and L.~Tavian  for useful discussions. Thanks also to the operators for their kind assistance from the CCC during all the measurements.

\newpage

\end{document}